\documentclass{article}
\usepackage[T1]{fontenc}

\usepackage{microtype}
\usepackage{graphicx}
\usepackage{subfigure}
\usepackage{booktabs} 
\usepackage{array} 
\usepackage{hyperref}

\usepackage[accepted]{icml2025}

\usepackage{amsmath}
\usepackage{amssymb}
\usepackage{mathtools}
\usepackage{amsthm}

\usepackage[capitalize,noabbrev]{cleveref}

\theoremstyle{plain}

\theoremstyle{definition}

\theoremstyle{remark}

\usepackage[textsize=tiny]{todonotes}
\usepackage{multirow}

\icmltitlerunning{CALMA: A Process for Deriving Context-aligned Axes for Language Model Alignment}

\begin{document}

\twocolumn[
\icmltitle{CALMA: A Process for Deriving Context-aligned \\Axes for Language Model Alignment}



\icmlsetsymbol{equal}{*}

\begin{icmlauthorlist}
\icmlauthor{Prajna Soni}{equal,sch}
\icmlauthor{Deepika Raman}{equal,sch}
\icmlauthor{Dylan Hadfield-Menell}{sch}
\end{icmlauthorlist}

\icmlaffiliation{sch}{Massachusetts Institute of Technology, Cambridge, MA, USA}

\icmlcorrespondingauthor{Prajna Soni}{prajna@mit.edu}
\icmlcorrespondingauthor{Deepika Raman}{deepikar@mit.edu}

\icmlkeywords{Machine Learning, ICML}

\vskip 0.3in]



\printAffiliationsAndNotice{\icmlEqualContribution} 

\begin{abstract}
Datasets play a central role in AI governance by enabling both evaluation (measuring capabilities) and alignment (enforcing values) along axes such as helpfulness, harmlessness, toxicity, quality, and more. However, most alignment and evaluation datasets depend on researcher‑defined or developer‑defined axes curated from non‑representative samples. As a result, developers typically benchmark models against broad (often Western‑centric) values that overlook the varied contexts of their real‑world deployment. Consequently, models trained on such proxies can fail to meet the specific needs and expectations of diverse user communities within these deployment contexts. To bridge this gap, we introduce CALMA (Context-aligned Axes for Language Model Alignment), a grounded, participatory methodology for eliciting context‑relevant axes for evaluation and alignment. In a pilot with two distinct communities, CALMA surfaced novel priorities that are absent from standard benchmarks. Our findings demonstrate the value of evaluation practices based on open‑ended and use‑case‑driven processes. Our work advances the development of pluralistic, transparent, and context-sensitive alignment pipelines.
\end{abstract}

\section{Introduction}
Value alignment ensures that LLMs comply with human values, ethical standards, and specified goals, and is typically achieved through reinforcement learning or supervised fine-tuning methods grounded in carefully curated datasets \cite{shen2023large}. However, current alignment datasets often rely on vague, researcher-defined axes (e.g., helpfulness, harmlessness, toxicity) and crowd-sourced preferences aggregated through platforms like MTurk. These axes are frequently shaped by a culturally homogeneous group of researchers \cite{casper2025pitfalls} and rarely reflect the multiplicity and diversity of real-world deployment contexts, thus creating a misalignment between datasets that serve as \textit{proxies} for human preferences and \textit{true human preferences}. Evidence has shown that this misalignment causes disparate model performance in varied and diverse contexts \cite{khandelwal2024indian}. 

While recent work has acknowledged the subjectivity of value alignment tasks \cite{chang2017revolt, prabhakaran2021releasing, bakker2022fine} and introduced participatory methods to gather diverse preferences \cite{Ganguli_Huang_Lovitt_Siddharth_Liao_Durmus_2023, kirk2024prism, kuo2024wikibench, bergman2024stela, meadows2024localvaluebench}, these approaches often remain limited by (1) researcher bias—where researchers inadvertently influence participants through priming, result interpretation, curated datasets, or predefined labels—and (2) poor user familiarity, where participants lack sufficient experience with the technology or process to offer informed feedback on their preferences.

To address these challenges, we propose CALMA (Context-Aligned Axes for Language Model Alignment), a non-prescriptive, open-ended, and participatory methodology guided by Grounded Theory \citep{charmaz2008constructionism}. CALMA elicits context-specific values from observed human interactions through four iterative stages: (1) Familiarize, (2) Interact, (3) Reflect, and (4) Discuss. 

We showcase and evaluate this approach through a pilot study with two distinct groups. Our study demonstrates a grounded and non-prescriptive approach to language model alignment that embraces subjectivity, community-specific definitions, and dialog-driven processes for context-specific alignment can help mitigate researcher bias. We conclude with a discussion of areas for future work. In particular, effective and engaging participant training is central to the success of an open-ended interpretive process like this. 

\section{Related Work}
\subsection{Technical Methods for Alignment}
\paragraph{General Alignment:}
Alignment methods for LLMs are broadly categorized into RLHF-based and SFT-based methods \cite{shen2023large}. \citet{glaese2022improving} introduce rule-conditional reward modeling, using natural language rules to guide RLHF. \citet{liu2022aligning} propose SENSEI, an Actor-Critic framework that embeds human value judgments into each generation step. Other RL-based methods leverage intermediate text edits \cite{liu2022second} or synthetic feedback \cite{kim2023aligning}. More recent techniques like Kahneman-Tversky Optimization (KTO) and Binary Classifier Optimization (BCO) align models using binary feedback on prompt-completion pairs \cite{ethayarajh2024kto, jung2024binary}.

\paragraph{Value-based Alignment:}
This growing subset includes methods like PALMS and Constitutional AI, which align models with specific values rather than general preferences. PALMS is an iterative process to change model behavior by crafting and fine-tuning on a dataset that reflects a predetermined set of target values \cite{solaiman2021process}, while Constitutional AI employs RLAIF guided by a set of principles referred to as the ‘Constitution’ to shape model behavior \cite{bai2022constitutional}. 

\paragraph{Multi-Dimensional Alignment:}
Emerging works move beyond a unidimensional axis of ``human preference'' by providing finer control over generated outputs and supporting context-specific evaluations and alignment \cite{sorensen2024roadmap}. FUDGE (Future Discriminators for Generation) explores descriptive, attribute-grounded conditioning to control text generation~\cite{yang2021fudge} while SteerLM applies attribute-conditioned SFT using explicitly defined traits like helpfulness, humor, and toxicity~\cite {dong2023steerlm}. \citet{wang2024arithmetic} propose Directional Preference Alignment (DPA), a multi-objective reward framework that enables intuitive, arithmetic control over LLM outputs by modeling user preferences as directions in a reward space.

\subsection{Participatory Methods for Alignment}
Recent work has also explored participatory approaches to identifying which values (axes) and preferences (positions along those axes) language models should align with.
\paragraph{Where along an axis?}
To address the diversity of human preferences, \citet{prabhakaran2021releasing} advocate for retaining annotator-level labels to preserve socio-cultural variation, rather than flattening them through aggregation. PRISM collects contextual feedback from 1,500 participants across varied demographics, illustrating divergent alignment norms across groups \citep{kirk2024prism} while Wikibench facilitates community-driven curation of evaluation datasets through deliberation to navigate consensus, uncertainty, and disagreement \citep{kuo2024wikibench}.

\paragraph{Which axes?}
Similar to CALMA's objectives, methods like Collective Constitutional AI~\cite{Ganguli_Huang_Lovitt_Siddharth_Liao_Durmus_2023} and STELA~\cite{bergman2024stela} depart from preference aggregation to elicit values from specific groups to inform alignment. STELA derives alignment rules from community evaluation of researcher-curated dialogue samples along pre-defined themes, while Anthropic similarly distills a ``constitution'' by analyzing public input (votes) gathered on Polis. \citet{weidinger2023using} apply social choice theory, showing that the “Veil of Ignorance” framework fosters fairness-oriented reasoning in value selection for alignment.

\section{Methodology}
\begin{figure}[ht]
    \begin{center}
    \includegraphics[width=0.6\linewidth]{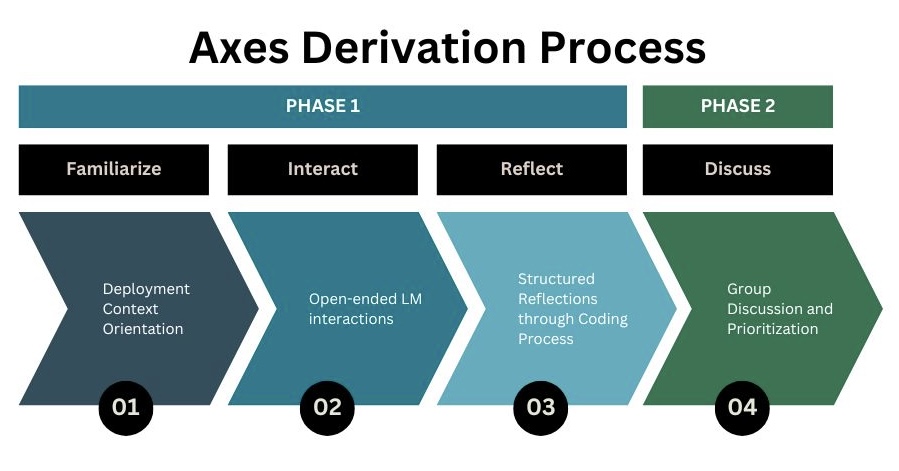}
    \end{center}
    \vspace{1em}
    \caption{The CALMA process}
    \label{fig:calma}
    \vspace{1em}
\end{figure}

CALMA applies the widely-used Grounded Theory \citep{charmaz2008constructionism} by centering interaction data in ontology creation and elicits context-specific axes through four distinct stages (Figure \ref{fig:calma} and details in \ref{sec:appendix}ppendix. The resulting axes help answer the question of which axes an LLM should be aligned along for a given context.

\paragraph{[1] \textit{Familiarize}: Deployment Context Orientation.}
To empower participants to provide meaningful input, the first stage builds participant familiarity with (1) the context within which the language model will be deployed, and (2) Grounded Theory application for the CALMA process. To minimize priming, participants were familiarized with the process using examples from an unrelated and orthogonal context.

\paragraph{[2] \textit{Interact}: Open-ended LM Interactions.}
To ground the value elicitation process, participants engaged in conversations with the model on topics relevant to its deployment context, generating a realistic and diverse interaction space for annotation. No limitations were placed on the manner of interaction.

\paragraph{[3] \textit{Reflect}: Structured Reflections through Open Coding.}
Participants analyzed their interactions through structured reflections guided by Charmaz’s grounded theory coding framework \citep{charmaz2008constructionism}. Using a custom tool, they annotated implicit and explicit values in their own interactions and a baseline set across all participants. Coding occurred in two stages: (1) Initial Coding, where participants labeled model behaviors with observed values, and (2) Focused Coding, where they refined and grouped related codes into final attribute clusters.

\paragraph{[4] \textbf{\textit{Discuss}}: Building Consensus as a Group}
Participants reconvened as a group to discuss their labels in a structured and open-ended format. The goal of the discussion is for participants to identify a set of most ``relevant'' attributes for the identified deployment context. 

\section{Findings from a CALMA Pilot}
We conducted a study with two different groups (MIT students and professionals from India) to derive context-specific axes for an ``Educational Assistant for History'' (see \ref{sec:appendix}ppendix for further details). This pilot study demonstrates CALMA’s ability to surface nuanced, non-prescriptive alignment axes rooted in participants’ lived contexts. In addition to analyzing the elicited outputs, we gathered feedback through surveys and interviews to evaluate the process itself. This section presents key findings that highlight the strengths, trade-offs, and challenges of adopting a grounded and participatory approach to language model alignment.  See Table~\ref{tab:mit_india_eval} , \ref{tab:llm_axes1}  and \ref{tab:llm_axes2} in the Appendix \ref{sec:appendix} for the final outputs from the CALMA pilot, which illustrate the observations in this section.

\subsection{Open Yet Contextual Model Interactions}
The open-ended design of the study allowed participants to explore a broad array of topics through interactions with the educational assistant, ensuring a diverse annotation space rooted in participant diversity and reflecting realistic conversations. While all participants annotated a small, shared set of baseline interactions to enable cross-participant comparison, they curated the remainder of their data through unique, self-directed interactions. Some participants chose to explore a single topic in depth, while others navigated multiple subjects shaped by their knowledge, ideologies, or curiosity. MIT participants discussed topics such as the assassinations of JFK and Malcolm X, the history of the American South, and the Middle East. In contrast, the Indian group focused on topics like ancient civilizations, colonial rule, and Kashmir.

\subsection{Articulating Values and Axes}
The unfamiliar task of articulating axes within the CALMA framework encouraged participants to think reflectively and creatively, resulting in varied outcomes shaped by their individual experiences and cultural contexts. For example, all MIT participants coded the following baseline model response:
\begin{quote}
    \textbf{LLM: }“\textit{...Thanksgiving is a holiday celebrated in the United States, traditionally on the fourth Thursday of November. It originated as a harvest festival and has been celebrated in the US since the early 17th century. The holiday is centered around a feast, typically featuring roasted turkey, mashed potatoes, stuffing, cranberry sauce, and pumpkin pie. The tradition is traced back to a 1621 celebration at the Plymouth Colony, where the Pilgrims, early European colonizers of the US, shared a meal with the Wampanoag Native American tribe to mark the harvest season.}”
\end{quote}
Participant annotations of this excerpt ranged widely—from `neutral but nuanced,' `factual,' and `dry' to `incomplete,' `racially biased,' and `American viewpoint.' These diverse interpretations underscore the effectiveness of CALMA’s open-ended design in capturing a plurality of perspectives without researcher priming.

Table \ref{tab:annotations} illustrates the annotation diversity across both groups and coding stages. Each group provided distinct emphasis that reflected their particular context. For example, Indian professionals emphasized religious and colonial bias, while the MIT group highlighted slavery, Indigenous narratives, and cultural framing.

\begin{table*}[t]
\caption{Annotation examples from MIT and India participants, organized by label type}
\label{tab:annotations}
\vskip 0.15in
\begin{center}
\begin{small}
\begin{sc}
\begin{tabular}{%
  p{0.05\textwidth}%
  p{0.15\textwidth}%
  p{0.80\textwidth}%
}
\toprule
\textbf{Group} & \textbf{Label type} & \textbf{Example annotations} \\
\midrule
\multirow{2}{*}{MIT} 
  & Open coding annotation 
  & “bias”, “biased”, “colonial bias”, “bias to American”, “bias to Britain”, “bias to British government”, “racially biased”, “religious bias”, “bias to slavery”, “bias to natives” \\
\cmidrule(lr){2-3}
  & Focused grouping label 
  & “biased”, “colonial bias”, “helpful bias”, “unhelpful bias” \\
\midrule
\multirow{2}{*}{India} 
  & Open coding annotation 
  & “bias”, “biased”, “western bias”, “authority bias”, “colonial bias”, “potential religious bias”, “US bias” \\
\cmidrule(lr){2-3}
  & Focused grouping label 
  & “bias”, “bias perspective”, “stereotype”, “marginalisation”, “religious bias”, “violence”, “bias: historical”, “bias: western”, “western bias”, “colonial bias”, “controversy” \\
\bottomrule
\end{tabular}
\end{sc}
\end{small}
\end{center}
\vskip -0.1in
\end{table*}

\paragraph{\textbf{Subjective Interpretations:}} 
A challenge in participatory alignment processes is the risk of researcher bias when interpreting subjective annotations. Both groups used the broad label `bias,' yet their interpretations and the targets in question varied: the Indian group emphasized religious and colonial bias, while the MIT group focused on racial and cultural framing (as summarized in Table~\ref{tab:annotations}).

Relying on aggregated labels without contextual grounding can obscure important distinctions and reintroduce researcher bias. To mitigate this, an example-driven and dialogic process in which participants refine and define their labels is essential for eliciting more accurate and actionable preferences (for example, see Table~\ref{tab:llm_axes2}).


\subsection{Uncovering Nuances with Intersectional Axes}
\paragraph{Intersectional Axes:}
Axes are not always orthogonal or independent; preferences along one axis often correlate with those on another. Participant discussions revealed the subjective nature of distinguishing correlated axes, such as ‘factuality’ and ‘citations,’ further highlighting the need to minimize researcher bias.
\begin{quote}
    \textit{``.. should we [...] combine fact and citation? Any fact claimed should have some citation, right?''} 
\end{quote}
\begin{quote}
    \textit{``to understand not just how credible a source is, but if there is a political leaning [...] indicating the different sources of power [...] so then we kind of cover all of that under fact.''} 
\end{quote}
\begin{quote}
    \textit{``For students, I think it’s very important to understand what ideology a certain interpretation is coming from […] so maybe fact-citation could be combined into one thing. And then the school of thought could be [a separate attribute]..."}
\end{quote}
Clearly outlining attribute definitions with examples enriches the alignment exercise by minimizing subjective interpretations and enhancing the clarity of axes distinctions.

\paragraph{Uncovering Nuances:}
The minimally moderated, open-ended sessions created space for participants to clarify the study’s goals and yielded rich insights into the complexities of consensus-building, particularly highlighting the \textbf{importance of deeper reflection and articulating values beyond simple ranking or aggregation schemes}.

Participants distinguished between concepts like values (which axes) and a preference (where along an axis), questioned the relevance of specific attributes in varying contexts, and debated whether certain factors such as local versus global geographical considerations should matter for the given deployment context. These discussions surfaced reflections such as:
\begin{quote}
    \textit{``...I learned Gutenberg invented the printing press in school but later learned that there were precursors in Korean and Chinese printing technology years before that..."}
\end{quote}
\begin{quote}
    \textit{``...if I'm a US high school student and I ask about the first President, obviously tell me George Washington. Right? But I don't know how to distinguish when you ask something about who invented printing where it's kind of a universal thing... [I want the model to] tell me about printing outside of the Western context [as well]..."}
\end{quote}
Such reflections highlight that participant disagreements observed in dialogic interactions surface underlying values and contextual assumptions. By surfacing these moments, the axis-derivation process becomes more than a data collection exercise, contributing to a richer and nuanced understanding of derived axes.

\section{Key Takeaways}
\begin{enumerate}
    \item \textbf{Capturing nuance requires non-prescriptive processes}: A user-led approach to prompting and annotation enables a rich and expressive label space to emerge, reflecting the subjectivity of value-related tasks and preserving the diversity of participant perspectives.
    \item \textbf{Dialogue supports the subjectivity inherent in non-prescriptive processes}: The range of annotations produced through grounded methods is best interpreted through dialogue rather than automated clustering or majority voting, allowing for clarification and deeper articulation of attribute meanings across perspectives.
    \item \textbf{Context-specific definitions and grounded examples enhance axis development}: The subjective nature of value labeling is contextualized through collaboratively defined terms and examples, anchoring axes in the lived experiences and priorities of specific contexts.
\end{enumerate}

\section{Limitations and Future Work}
Scaling CALMA across diverse contexts and communities will require iterative adaptation and improvements at different steps, some of which we illustrate below based on our learnings from the pilot user studies: 

\paragraph{\textbf{Before CALMA/ Participatory Considerations   }}
\begin{itemize}
    \item \textbf{Training Participants for Grounded Theory:} Engaging in such a specialized and time-intensive process should be preceded by necessary training along with an element of testing to ensure a baseline level of understanding of the tasks. Value elicitation and the application of Grounded Theory Coding are not familiar or intuitive exercises, and inadequate training can lead to generic or flawed descriptive annotations. For instance, `summary', `introduction', 'conclusion' were labels we initially observed in our pilot which evolved to more nuanced value-based labels as we updated our training. Furthermore, increasing participant familiarity by contextualizing the study tasks into familiar applications and tangible use cases (e.g., ChatGPT) can help reduce task ambiguity.
    \item \textbf{Participant Selection:} As our study was intended as a pilot, participant groups were not sampled to be representative. In future applications, use-case-driven participant selection will be crucial. While we do not prescribe a specific definition of \textit{context} or \textit{community}, the process is adaptable to varying definitions and scales. Drawing on HCI best practices, recruitment should be grounded in the application's intended context (e.g., teachers, students, and civil society actors for the presented pilot).  Including historically marginalized groups is particularly important for uncovering overlooked values. For instance, as educators in under-resourced schools may surface concerns quite different from those in affluent districts.
    \end{itemize}

\paragraph{\textbf{Scaling CALMA/ Revisions to Phase One   }}
\begin{itemize}
\item \textbf{Automating Label Clustering:} To automate clustering of participant labels while preserving CALMA’s non-prescriptive nature, future iterations could increase content overlap. Increasing the overlap in the themes explored, as well as making participants annotate a mix of their own and others’ responses can expand the diversity of labeled content. 
\item \textbf{Processing outputs from }\textit{\textbf{Reflect}}\textbf{:} Alternatives to manual mapping could involve training classifiers on user labels, clustering responses using contextual embeddings, or applying active learning within gamified platforms.

\end{itemize}

\paragraph{\textbf{Refining CALMA/ Revisions to Phase Two}}
\begin{itemize}
\item \textbf{Ensuring Inclusion in Group Discussions:} Group settings can reproduce societal power dynamics, risking the exclusion of subaltern perspectives. To mitigate this, smaller curated sessions can be used to amplify specific experiences, and domain experts (e.g., researchers or advocates) may be brought in to help center marginalized views.
\item \textbf{Re-imagining Group Discussions:} Scaling to larger and more representative samples may require new deliberation formats. Prior work showcasing synchronous argumentation (e.g., Cicero \citep{chen2019cicero}), structured disagreement (e.g., WikiBench \citep{kuo2024wikibench}), and consensus-driven platforms (e.g., Polis \citep{small2021polis}) offer scalable alternatives. Supplementary techniques such as aggregation, voting, and social choice theory help balance consensus-building with preserving diversity \citep{fish2023generative}. 
        \end{itemize}

\paragraph{\textbf{After CALMA/}
 Context-specific Alignment and Evaluations}
 \begin{itemize}
\item CALMA surfaces alignment axes with definitions and real interaction examples, which can be leveraged to create training and preference datasets tailored to specific use cases. These datasets can support alignment and evaluation pipelines, can be used to inform annotation guidelines for identifying attribute-relevant content in model outputs, and facilitate the development of context-specific LLM judges \citep{zheng2023t}. They can also be used in alignment techniques such as SteerLM \citep{dong2023steerlm} and RLHF. 
\item For \textbf{scalable deployment}, few-shot prompts can be refined with user definitions and interaction examples, and lightweight classifiers can be trained to identify attribute-relevant content in model outputs. These classifiers can enable in-context learning setups for semi-automated dataset expansion. 
\item To assess the downstream impact, interaction data and annotations from CALMA can be used to compare models trained along CALMA-identified preferences with those trained on traditional datasets. A/B testing among participants from the same demographic can allow for early evaluation of model preference within a specific application context.
\end{itemize}
\section{Conclusion}

This paper introduces the CALMA process—a grounded, non-prescriptive, and participatory methodology for deriving community-relevant axes in language model alignment. CALMA reduces researcher bias and surfaces rich, contextually grounded values and perspectives. We evaluated its viability through a proof-of-concept application of CALMA with two populations. In comparison to prescriptive labeling schemes, CALMA leverages dialogue and community grounding to produce nuanced evaluations and targets for AI alignment. Our work demonstrates how alignment pipelines can better incorporate pluralism, transparency, and contextual sensitivity. 

\section*{Impact Statement}
This paper presents a participatory methodology for aligning language models with the contextual values of specific communities. By centering stakeholder perspectives in the derivation of value-relevant axes, the CALMA approach aims to reduce misalignment between model behavior and the expectations of affected populations. While this work contributes to the broader goal of socially informed and value-sensitive AI, it also raises important ethical considerations around representation, power dynamics in participatory processes, and the potential misuse of community-derived preferences. We believe this method encourages more inclusive and accountable AI development but emphasize the need for careful deployment to avoid reinforcing dominant perspectives or excluding marginalized voices.

\bibliography{example_paper}
\bibliographystyle{icml2025}

\newpage
\appendix
\onecolumn
\section{Appendix} \label{sec:appendix}
\subsection{Additional Details on the CALMA Process}

\subsection{CALMA Tool Details}

The CALMA methodology was tested by using a platform adapted from the Open Coding for Machine Learning tool\footnote{https://github.com/Algorithmic-Alignment-Lab/OpenCodingForMachineLearning/tree/community\_ocml} \cite{price2022open}. The language model interface for the \textit{Interact} step was built using Llama-70B \& Mistral-8x7B, where the following system prompt was used to contextualize the model responses: \textit{``You are an education assistant for high school students studying history in India. Answer questions to help pique the student's curiosity of the topic. Limit your answers to 2 paragraphs."} 
To maintain a non-prescriptive format and center participant agency, the design of the \textit{Interact} phase was informed by insights from participant interviews conducted at the end of the \textit{Reflect} phase.
\subsection{Group Discussion Design: \textit{Interact}}
\begin{enumerate}
    \item \textbf{Session Artifacts}: Before the group discussion, three artifacts were shared with each participant: (i) their respective interactions, annotations, and labels; (ii) a summary document with topics the group explored and a word cloud capturing the frequency of the group’ annotations based on a simple word count, and (iii) an optional-to-view video that contextualized the outputs of the study in the language model alignment pipeline.
    \item \textbf{Session Orientation:} Researchers introduced the objectives and tentative structure of the session but kept their guidance minimal, encouraging an atmosphere of open-minded and respectful dialogue to platform multiple perspectives. 
    \item \textbf{Segment One: Initial Individual Ranking of Attributes}: Each participant identified and defined their top three attributes after reviewing their Phase One interactions. This process helped them reacquaint with the interactions and independently form preferences before engaging in discussion. Each attribute's definition was then added to the Miro board for reference throughout the discussion. 
    \item \textbf{Segment Two: Exploring Embedding Space}: Participants then explored a visual embedding space on the Miro interface\footnote{Miro is a digital collaboration platform designed to facilitate remote and distributed team communication and project management.}, created by researchers to arrange labels (accompanied by annotations) based on their semantic meaning, similar to affinity mapping \cite{rodighiero2018mapping}. For example, ‘balanced,’ ‘diplomatic,’ and ‘nuanced’ would appear close to each other in the embedding space, similarly ‘biased’ and ‘one-sided’ were close. This format aimed to present the group's collective interaction data in a more comprehensible format while minimizing any hierarchy within the elicited labels. 
    \item \textbf{Segment Three: Individual Presentations}: Each participant presented their top three attributes to the group, explaining their interpretation of that value as observed in their interactions with the model. This served as an icebreaker and an opportunity for everyone to contribute to the discussion.
    \item \textbf{Segment Four: Group Discussion}: Participants discussed their perceptions of the task, the values they coded, and any reactions or clarifications regarding the labels on the board. Through continued dialogue, they collectively identified axes pertinent to their community, defined the final set of attributes, and supplemented them with examples of model interactions. The group then ranked their top three and five attributes from this set.
    \item \textbf{Segment Five: Individual Ranking of Attributes}: After segment four, each participant independently rated their agreement with the group's attribute definitions on a five-point Likert scale. They also submitted their individual rankings of the attributes they found most pertinent. Comparing Segments 1 and 5 helped illustrate the impact of the group discussion in building consensus and influencing individual perceptions of the attributes.
\end{enumerate}

\subsection{Pilot Study}

The CALMA process was evaluated by conducting user studies with two distinct groups. The deployment context that was introduced to these groups in order to derive community-relevant axes, was an LLM adapted as a ”History Educational Assistant” for high school students. We chose History as it varies contextually and is arguably subjective to an extent. All study protocols were approved by MIT’s COUHES and conducted in compliance with relevant regulations and ethical norms. 

The first pilot group consisted of 11 MIT students (US citizens), while the second, in-context group included 15 working professionals from India. Both groups were proficient in English and had basic exposure to language models. Participants provided informed consent and were compensated USD 17 per hour for their time. All sessions for Phase One were conducted online via Zoom. Phase 2 was conducted in person for the MIT group and online for the Indian participants.

Since the premise of this study was to test the grounded and non-prescriptive CALMA process, the groups were not sampled to be representative or act as a set of experts representing a given community, so the final attributes cannot be ascribed as indicative of any community’s preferences. 



\subsubsection{Notable Insights: }
Despite the rich labels we gathered through our cases studies, important insights that could help strengthen the \textit{Familiarize and Discuss} steps of the CALMA process were:
\begin{itemize}
    \item \textbf{Split on “Consensus”}: Interview feedback indicated that participants were divided on the consensus-building aspect of the discussion. MIT participants found it conducive to reaching consensus, whereas Indian participants were split—some citing limited discussion time as a constraint, and others acknowledging their own inflexibility as the main issue.
    \item \textbf{\textit{On} Axes vs \textit{Where} Along the Axes}: The discussions also highlighted the challenge of depoliticizing such a process and aligning personal values with the group’s consensus, as illustrated by this participant's experience:
    \begin{quote}
    \textit{“While I was relieved that we weren't attempting to find political consensus with a group of strangers, I found it difficult to depoliticize this process for myself. Because I think it is precisely where we stand along the dimension that makes a value/quality apparent to us.”}
    \end{quote}
    \item \textbf{Group Size and Discussion Format:} Owing to scheduling conflicts, the Indian group had 12 participants joining virtually whereas the MIT group had only 6. The smaller size of the MIT group allowed for more discussion time to furnish detailed attribute definitions, and the in-person format benefited from non-verbal cues and allowed for a more organic sense of ‘community’ to emerge in the discussion. A shortcoming of both groups however, was the over-representation of male-identifying participants. 
\end{itemize}

These insights provided important lessons on the types of contextual clarity and group curation that can further empower participants in future iterations of this process to enable their meaningful participation. 

\subsection{Comparison Tables}

Note: As mentioned in the \textit{Notable Insights} section, the final axes in  (for example, see Table~\ref{tab:mit_india_eval}) are influenced by some of the logistical realities of running such a participatory study. The Indian group lost time to technical difficulties in the virtual format and its larger size resulted in a paucity of time for comprehensive discussion. The labels have also been reordered in this table for easier comparison between the groups. 

\vspace{-0.3em}
\begin{table}[htbp]
\caption{Derived Axes from MIT and India Pilot Studies}
\label{tab:mit_india_eval}
\vspace{-0.6em}
\begin{center}
\scriptsize  
\setlength{\tabcolsep}{2.5pt}  
\renewcommand{\arraystretch}{2}  
\begin{tabular}{%
p{0.15\textwidth}
p{0.33\textwidth}
p{0.15\textwidth}
p{0.33\textwidth}%
}
\toprule
MIT & Definition & India & Definition \\
\midrule
Cultural Context& The degree to which the model returns simple facts versus returning facts along with cultural context to see how different groups were impacted by historical events; The LLM is able to distinguish and clarify minority versus majority perspectives on historical events.& Inclusive& The LLM should be able to pull up examples and situations that are inclusive to a larger set of audience or a specific set of audience, and to include different key perspectives and schools of thought (e.g. coloniser and colonised).\\

Source & Provide a diversity of sources, or specify the origin of source for the response given.& Fact / Power& The response makes an accurate factual claim, cited from the source and indicating an alignment of differing sources of power.\\

Empathy& The degree to which the LM uses emotional or empathetic language to show understanding of the user's emotions and solidarity with the user's community; Not just presenting factual information, but also expressing an emotional connection to the user's experience; Provides context from an empathetic perspective that comprehensively captures attitudes towards historical events.& Emotion& The LLM should respond in ways that are compassionate, kind and empathetic both towards the user and the subject it is talking about.\\

Comprehension Level& Adjust complexity of responses to meet reading comprehension level of the user.& Complexity& Level of complexity of the answer: a spectrum of levels can be used for the same response\\

Localization/

Geographic Breadth& The geographic/regional scope of the responses for which the LM are presented; Should the LM present information that is only relevant based on a user's location or should it provide interregional/ global context to the user based on the question?& Specificity& Response is specific and concrete, both with respect to concepts and examples. It shouldn't be vague or incomplete and cover all key aspects of the matter at hand.\\

Confidence Level& The model is able to admit its lack of info, include disclaimers, acknowledge limitations in breadth \& depth of information available.& Misleading& The LLM should be controlled on sharing on being reductive at times and over simplifying things because sometimes this leads to misinformation.\\

— & — & Prescriptive& The LLM should not be overly prescriptive/pedantic in its responses; it should not ask a particular way of thinking/acting of the user.\\

— & — & Schools of Thought& The LLM response is based on a certain school of thought (without necessarily revealing it). Modifying this is a potentially useful thing for educators to do to enable students to explore different interpretations of facts.\\

— & — & Completeness& Response covers all key aspects of the matter at hand.\\

— & — & Citation& Specific sources the answer drawn from, response covers all key sites the resources have been taken.\\

— & — & Wrong Premise/ 

Inaccurate Information& Wrong information, the LLM should ideally be able to catch a wrong premise and oppose it, versus assuming it to be true and going ahead with it.\\
\bottomrule
\end{tabular}
\end{center}
\vspace{-1em}
\end{table}

\begin{table}[ht]
\caption{Select Axes with Sample Interactions Tagged by MIT Participants}
\label{tab:llm_axes1}
\vskip 0.15in
\begin{center}
\begin{small}
\begin{tabular}{p{0.15\textwidth}p{0.75\textwidth}}
\toprule
\multicolumn{2}{l}{Label: \textbf{Cultural Context}} \\
\midrule
\multicolumn{2}{p{0.9\textwidth}}{Label Definition: The degree to which the model returns simple facts versus returning facts along with cultural context to see how different groups were impacted by historical events; the LLM is able to distinguish and clarify minority versus majority perspectives on historical events.} \\
\midrule
\textit{Sample Interaction 1}& The exact number of people who died during the Trail of Tears is not known, but it is estimated that between 4,000 and 8,000 Native Americans died during the removal. This represents about one-quarter to one-third of the total population of those removed. The deaths were primarily due to disease, starvation, and exposure during the long journey. The Trail of Tears was a tragic and painful chapter in US history, and serves as a reminder of the devastating impact of forced relocation and the violation of Native American sovereignty. \\
\textit{Sample Interaction 2}& The Cherokee were forced to adopt European-American farming practices, which led to a decline in their traditional agriculture and foodways. This process of cultural assimilation continued well into the 20th century, and it had a profound impact on the Cherokee Nation's sense of identity and self-worth. \\
\midrule
\multicolumn{2}{l}{Label: \textbf{Source}} \\
\midrule
\multicolumn{2}{p{0.9\textwidth}}{Label Definition: Provide a diversity of sources, or specify the origin of the source for the response given.} \\
\midrule
\textit{Sample Interaction 1}& The term "colonialism" was first used in the 16th century to describe the process of establishing and maintaining colonies in the New World. The term was used by European powers such as Spain, Portugal, France, and England, who established colonies in the Americas, Africa, and Asia. The concept of colonialism was further developed and theorised by scholars and philosophers such as Immanuel Kant and Adam Smith in the 18th and 19th centuries. \\
\textit{Sample Interaction 2}& The term "terrorism" has a more complex and contested history. The term was first used in the 18th century to describe acts of violence and intimidation committed by the French Revolutionary government against its opponents. It was later used to describe acts of violence and terror committed by various political and social groups in the 19th and 20th centuries, including anarchists, communists, and nationalist movements. \\
\bottomrule
\end{tabular}
\end{small}
\end{center}
\vskip -0.1in
\end{table}
\begin{table}[ht]
\caption{Select Axes with Sample Interactions Tagged by India Participants}
\label{tab:llm_axes2}
\vskip 0.15in
\begin{center}
\begin{small}
\begin{tabular}{p{0.15\textwidth}p{0.75\textwidth}}
\toprule
\multicolumn{2}{l}{Label: \textbf{Cultural Context}} \\
\midrule
\multicolumn{2}{p{0.9\textwidth}}{Label Definition: The LLM response is based on a certain school of thought (without necessarily revealing it). Modifying this is a potentially useful thing for educators to do to enable students to explore different interpretations of facts.} \\
\midrule
\textit{Sample Interaction 1}& Certainly! The term "Bengal Frontier" can refer to different geographical areas and time periods, and historians and scholars may use varying definitions and scopes depending on their research focus. Here are a few examples of how scholars have defined and studied the Bengal Frontier in their works: [examples]\\
\textit{Sample Interaction 2}& However, the Out of India hypothesis has been widely criticized by mainstream scholars of linguistics, archaeology, and anthropology. The majority of scholars reject the hypothesis, citing a lack of evidence and methodological flaws in the arguments put forward by its proponents. The Kurgan hypothesis remains the dominant theory for the origins of the PIE language.\\
\midrule
\multicolumn{2}{l}{Label: \textbf{Fact/Power}} \\
\midrule
\multicolumn{2}{p{0.9\textwidth}}{The response makes an accurate factual claim, cited from the source and indicating an alignment of differing sources of power} \\
\midrule
\textit{Sample Interaction 1}& The impact of the British Empire on India is a complex and widely debated topic. On one hand, the British Empire introduced modern education, legal and administrative systems, and infrastructure such as railways, telegraph and postal services. These developments have contributed to India's growth and development in many ways.On the other hand, the British Empire also exploited India's resources, imposed heavy taxes, and caused widespread poverty and famines. The British policies led to the destruction of traditional Indian industries and the impoverishment of the Indian peasantry. The British Empire also enforced policies that divided the Indian people and led to the partition of India in 1947, resulting in widespread violence and displacement. Therefore, whether the British Empire was a good or bad thing for India depends on the perspective and the specific aspects of its impact that one chooses to focus on.\\
\textit{Sample Interaction 2}& The New York Times, like many other news organizations, has editorial guidelines and policies that influence how they report on events and issues. While the newspaper is committed to presenting accurate and balanced reporting, its editorial decisions may sometimes be perceived as favoring one side over another or avoiding certain controversial topics. In the context of the Israeli-Palestinian conflict, the New York Times has been criticized for not directly attributing responsibility for Palestinian deaths to Israeli actions, which some argue amounts to self-censorship or bias.\\
\bottomrule
\end{tabular}
\end{small}
\end{center}
\vskip -0.1in
\end{table}


\end{document}